\documentclass[aps,pra,preprint,groupedaddress]{revtex4-1}
\usepackage{amssymb}
\usepackage{graphicx}

\begin{document}

\title{Landau Levels in Lattices with Long Range Hopping}

\author{Hakan Ataki\c{s}i}
\affiliation{Department of Physics, Bilkent University, 06800, Ankara,TURKEY}

\author{M. \"{O}. Oktel}
\email[]{oktel@fen.bilkent.edu.tr}
\affiliation{Department of Physics, Bilkent University, 06800, Ankara,TURKEY}

\date{\today}

\begin{abstract}
In the presence of a periodic potential Landau levels (LLs) are broadened, forming a barrier for accurate simulation of fractional quantum Hall effect using cold atoms in optical lattices. Recently, it has been shown that the degeneracy of the lowest Landau level (LLL) can be restored in a tight binding lattice, if a particular form of long range  hopping is introduced [E. Kapit and E.~J. Mueller, Phys. Rev. Lett. 105, 215303 (2010)]. In this paper, we investigate three problems related to such quantum Hall parent Hamiltonians in lattices. First, we show that there are infinitely many long range hopping models in which a massively degenerate manifold is formed by lattice discretizations of wavefunctions in the continuum LLL. We then give a general method to construct such models, which is applicable to not only the LLL but also higher LLs. We use this method to give an analytic expression for the hoppings that restore the LLL, and an integral expression for the next LL. We also consider whether the space spanned by discretized LL wavefunctions is as large as the space spanned by continuum wavefunctions and find the constraints on magnetic field for this condition to be satisfied. Finally, using these constraints and first Chern numbers, we identify the bands of the Hofstadter butterfly that correspond to continuum LLs. 
\end{abstract}

\pacs{}

\maketitle

\section{Introduction}
\label{sec:Introduction}

Cold atom experiments are ideal for controlled exploration of some of the fundamental problems in many particle physics. To this date, there has been great progress for some models like resonant interactions \cite{resonant} and Mott transition \cite{luttinger}, however one of the most important areas, fractional quantum Hall effect, has remained elusive. Simulating quantum Hall effects by rotation \cite{rotation} and artificial magnetic fields \cite{artificial} in harmonic traps has not been possible as these systems require extremely fine control over the uniformity of the artificial field or the rotation rate \cite{reviewCooper}. Recently, it has been demonstrated that optical lattices offer a variety of options for creation of high artificial magnetic fields \cite{latticemagnetic}. Optical lattice systems are generally easier to control and the filling factor, the ratio of number of particles to number of flux quanta, can be lowered by standard methods in these systems \cite{opticallattice}. It is reasonable to expect that quantum Hall physics can be probed by optical lattice experiments in near future.

\begin{figure}[ht]
\includegraphics[scale=0.6]{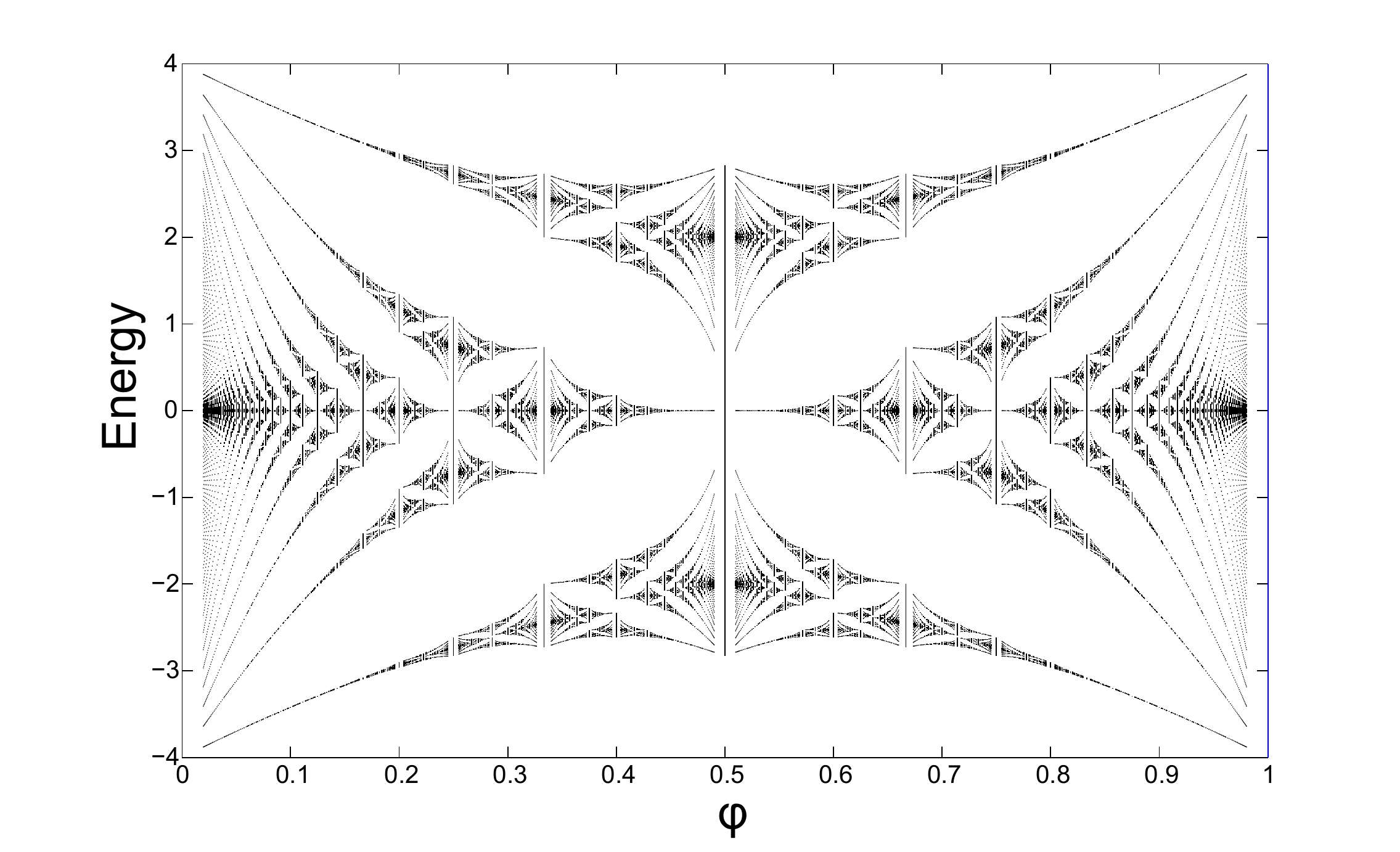}
\caption{\label{fig:Hofstadter} Energy bands for nearest neighbor tight binding model on a square lattice \cite{butterfly}. Energy is measured in units of nearest neighbor hopping and $\phi$ is the ratio of magnetic flux per plaquette to flux quantum. Energy bands are displayed for all fractions $\phi=p/q$ with $q\le80$.}
\end{figure}

 However, the presence of the optical lattice potential changes the physics of the quantum Hall problem significantly. A periodic potential lifts the degeneracy of the LL and broadens them into bands. The number of magnetic bands and their widths depend critically on the lattice constant \cite{bands}. The resulting energy spectrum is generally a self similar fractal, for example a nearest neighbor tight binding model in a square lattice gives the Hofstadter Butterfly \cite{butterfly} spectrum.(See Fig \ref{fig:Hofstadter}.) The broadening of the LLs is especially important for the many particle problem. Unless the LLs are flat enough, kinetic energy will dominate over interactions and highly correlated states will not be observed. Even when the LL broadening is made much smaller than the interaction energy scale, physics of the lattice system may differ from that of the continuum. There is compelling evidence for the existence of fractional quantum Hall states on the lattice which lack a continuum limit\cite{latticeFQH}. Thus, it would seem that an optical lattice experiment which has energy bands and wavefunctions with discrete symmetries will not directly probe the usual quantum Hall physics.

Cold atom systems are extremely versatile and in a surprising development Kapit and Mueller has shown that by introducing a particular form of long range hopping to the square lattice it is possible to create a system that exactly mimics the LLL on the lattice \cite{kmu10}. In a lattice with Kapit-Mueller hoppings there is an infinitely degenerate manifold, and the states forming this manifold are exactly LLL wavefunctions sampled on the square lattice. Thus, such a lattice Hamiltonian forms a parent Hamiltonian for certain fractional quantum Hall states. For example, the Laughlin wavefunction for $\nu=1/3$ is an exact eigenstate on this lattice, and it is the ground state in the presence of on-site repulsive interactions. 

The importance of this long range hopping model is twofold. First, hopping strengths in an optical lattice can be modified by tailoring the lattice potential or by dynamical hopping methods \cite{opticallattice}, and an optical lattice with long range hopping can experimentally realize fractional quantum Hall states. Second, as a theoretical development, this model bridges the gap between continuum and lattice models for quantum Hall physics, and makes it possible to simulate quantum Hall effects with a new method \cite{kgm12}. Hence, it is important to investigate and generalize this long range hopping model, both to understand the underlying physics in greater detail and to search for models that can be experimentally implemented. 

Beyond cold atom systems there are recent proposals for probing lattice quantum Hall physics using circuit QED \cite{girvin,hafezi}, coupled dissipative optical resonators \cite{onur} or in a general quantum circuit architecture \cite{kapit}. While these systems take a more indirect route towards realization of bosonic quantum Hall states, they present great potential for realization of lattice Hamiltonians with longer range hopping. Models considered here can aid the design of such artificial bosonic lattices to probe correlated states.

In this paper, we consider the general properties of long range hopping Hamiltonians which can mimic LL structure accurately. First, we investigate if Kapit-Mueller Hamiltonian is unique in creating a LLL on the lattice. A simple argument shows that there are infinitely many models which recreate the LLL in the desired form. For this argument, we define the projection operators which separates the subspace of the Hilbert space that is spanned by lattice sampled LLL wavefunctions. While two different states in the continuum LLL are orthogonal, lattice wavefunctions that are formed by sampling them are not. Thus, explicit expression for the projection operators are obtained by utilizing the symmetries of the problem. 

Ref \cite{kmu10} relied on a lattice sum identity to find the long range hopping model. This identity is remarkable in its efficiency, however opaque in underlying physics and not suitable for generalization. Here, using the projection operators, we formally describe a method to construct the most general long range hopping model that has a LLL. The method can be used in an arbitrary lattice, and to reconstruct any desired LL. For the LLL, we start from the formal expression in terms of the projection operators and find analytic expressions for a sample long range hopping model which has two infinitely degenerate bands, one being the desired LLL on the lattice. For the first excited LL,  we calculate the required hopping strength between any two lattice sites in terms of an integral, which can be evaluated numerically. For both cases, we numerically diagonalize the resulting Hamiltonian and verify that the desired LL is flat to our numerical precision. 
In the models we give, the hopping strength between two sites decay exponentially with distance, and is negligible after the third nearest neighbors. Thus, although infinitely long range hopping is required for exactly flat bands, an experimental realization can obtain high degeneracy by facilitating hopping to the first few nearest neighbors. 

Our analysis also enables us to answer a related question: ``What is the largest lattice constant one can use to discretize space for a particle under a magnetic field and still describe its dynamics correctly?". We find that for a particle in the LLL, any lattice size is adequate as long as the flux per plaquette of the lattice, $\phi$, is below $1$. For a particle in the first excited LL the lattice has to be finer so that $\phi < \frac{1}{2}$. Our numerical results suggest that for the n$^{th}$ LL, the constraint is $\phi<\frac{1}{n+1}$.  

When long range hopping as described in this paper is not present, it is not clear how one can define Landau Levels on a lattice. Nonetheless, by investigating the Hofstadter butterfly for low magnetic fields, one can observe that the Landau Level structure is present and identify certain bands with Landau Levels \cite{latticeLL}. The continuity of the Butterfly spectrum withstanding, it is hard to extend this identification to beyond $\phi \approx 1/3$ \cite{sorensen}. We find that an unambiguous identification can be made by considering the discretized wavefunctions of a LL and asking which bands of the Hofstadter spectrum are formed by states lying entirely in the manifold spanned by these discretized states. A continuous connection between the LL set and the Hofstadter spectrum is obtained by turning off the hopping strengths except for nearest neighbors. We empirically find that first Chern numbers of the largest gaps facilitate quick identification of corresponding LL for any band.

The paper is arranged as follows. In the next section, we introduce the problem in detail, set the notation and review the result of Ref.\cite{kmu10}. Section \ref{sec:Projection} contains the details of the projection operators and the general expression for long range hopping models with the desired property. In Section \ref{sec:LLL}, the method in the previous section is used to calculate explicit expressions for hopping strengths that reconstruct the LLL. In Section \ref{sec:1LL}, we first generalize the calculation to first excited Landau Level (1LL) and then present the results related to identification of LLs in the lattice problem. We conclude in Section \ref{sec:Conclusion}  with a summary and a discussion of the implications of our results. The appendix contains detailed evaluation of the integral used to calculate the LLL Hamiltonian.

\section{Problem Definition}
\label{sec:Definition}

The Hamiltonian for a particle of mass $m$ and charge $e$ on a plane subject to a perpendicular magnetic field $B$ is
\begin{equation}
{\cal H}_{\rm cont} = \frac{1}{2 m} \left( \vec{P} - e \vec{A} \right)^2,
\end{equation}
where $P$ is the is the momentum operator and $\vec{A}$ is the vector potential. We choose the Landau gauge,
\begin{equation}
\vec{A} = B {\cal X} \hat{y},
\end{equation}
with ${\cal X},{\cal Y}$ the continuous coordinates of the plane.
The eigenfunctions are labeled by the LL index, an integer $n$, and a real number $k_y$. The wavefunctions in the LLL with $n=0$, 
\begin{equation}
\Psi_{k_y}({\cal X},{\cal Y})= \frac{1}{( \pi \ell^2)^{\frac{1}{4}}} e^{-\frac{({\cal X} - {\cal X}_0)^2}{2 \ell^2}} e^{i k_{y} {\cal Y}},
\end{equation}
are degenerate with energy $\epsilon_0= \hbar \omega_c$. Magnetic length $\ell=\sqrt{\frac{\hbar}{m \omega_c}}$ and cyclotron frequency $\omega_c = \frac{e B}{m} $, follow the standard definitions and ${\cal X}_0=\frac{\hbar k_y}{e B}$.

Now we introduce a lattice model, with one state localized at each site of the square lattice 
\begin{equation}
\vec{r}= x \hat{x} + y \hat{y},
\end{equation}
where the coordinates $x$ and $y$ take on integer values so that the lattice constant is unity. The localized state at $x,y$ is created over the vacuum with
\begin{equation}
|x,y\rangle = a^{\dagger}_{x,y} | {\mathrm vac} \rangle,
\end{equation}
and, any one particle wavefunction can be written as
\begin{equation}
|\psi\rangle  = \sum_{x,y} \psi(x,y) |x,y \rangle.
\end{equation}

The objective is to find a one particle Hamiltonian, defined through the hopping strengths $J$ as
\begin{equation}
{\cal H} = \sum_{x,y} \sum_{x',y'} J(\vec{r},\vec{r'}) a^\dagger_{\vec{r}} a_{\vec{r'}},
\end{equation}
so that  a massively degenerate manifold is formed by lattice states $\psi_{k_y}$, which are obtained by sampling the continuum LLL wavefunctions at lattice points
\begin{equation}
\psi_{k_y}(x,y)= \Psi_{k_y}({\cal X}=x,{\cal Y}=y).
\end{equation}
The long range hopping Hamiltonian can only depend on magnetic flux per plaquette of the lattice $\phi$. We will assume that $\phi=p/q$ is a rational number with co-prime integers $p$ and $q$.

If the sampled wavefunctions are in one to one correspondence with the continuum LLL wavefunctions, and are degenerate, this lattice Hamiltonian will accurately mimic the continuum model. On site interactions do not significantly modify the correspondence between the lattice and the continuum problems, especially for fractional quantum Hall states which  minimize the local interaction between particles subject to the constraint of Landau levels.

\begin{figure}[ht]
\includegraphics[scale=0.6]{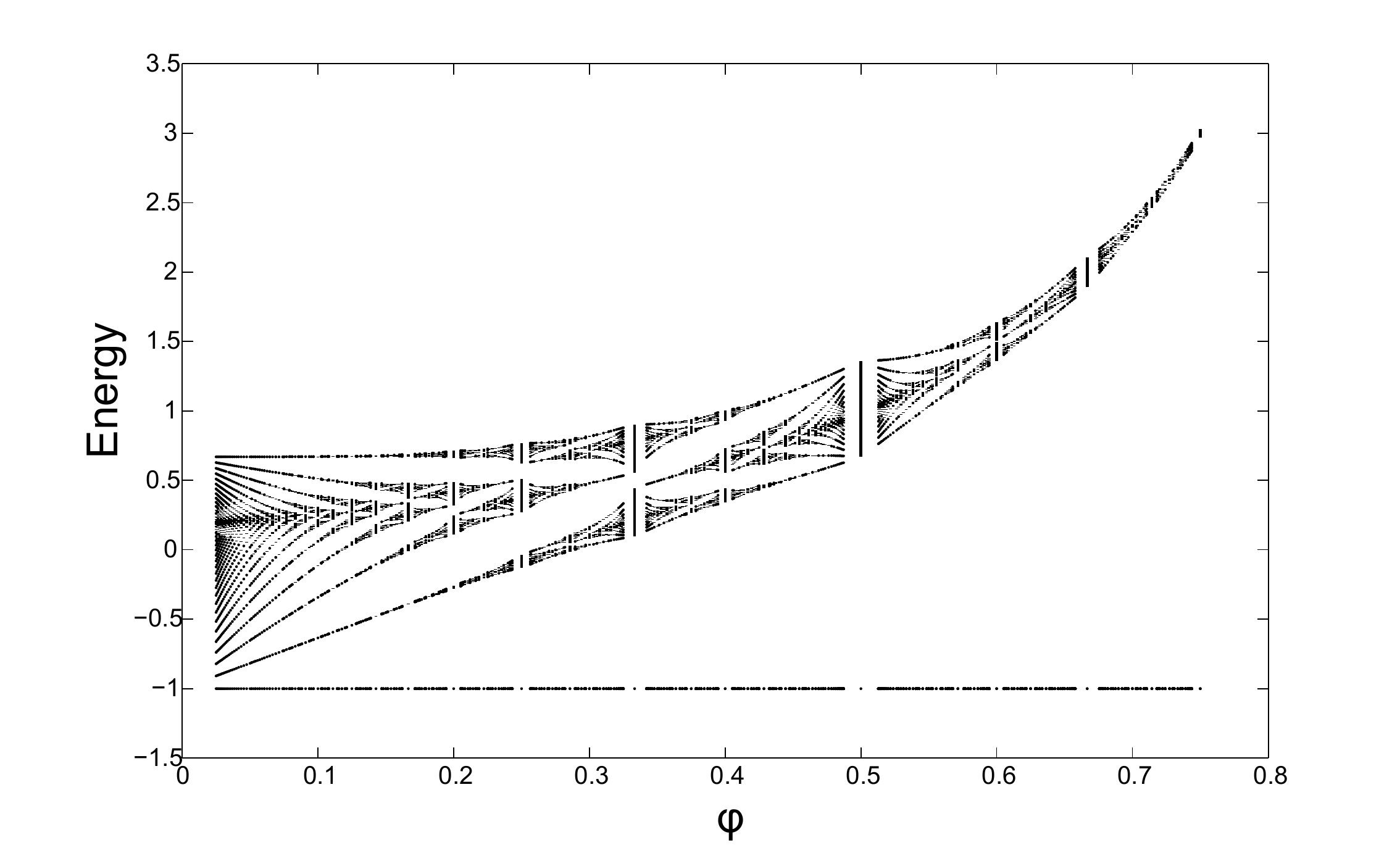}
\caption{\label{fig:KMButterfly} Energy bands for a square lattice with long range hopping defined in Ref. \cite{kmu10}, given in Eq.\ref{KMHopping}. Energy is measured in units of nearest neighbor hopping strength as in Fig.\ref{fig:Hofstadter}, while $\phi$ is flux quantum per plaquette of the lattice. For all values of $\phi$ there is a massively degenerate band with energy $-1$, while higher bands are broad with $\phi$ dependent energy. The bands are displayed for $\phi<.75$ to visualize the structure of the higher bands. }
\end{figure}

It is of course not immediately clear that such a lattice model can be found. The simplest hopping models, such as the Hofstadter Hamiltonian with
\begin{equation}
J(\vec{r},\vec{r'})= \left\{ \begin{array}{r@{\quad : \quad}l} -e^{i \pi \phi (y-y')(x+x')} & |\vec{r}-\vec{r}|=1 , \\ 0 & \mathrm{otherwise} , \end{array} \right.
\end{equation}
hopping only to the nearest neighbors, gives a spectrum of bands which are critically dependent on the flux per plaquette (See Fig. \ref{fig:Hofstadter}). The first example of a long range hopping model with a LLL was given by Kapit and Mueller \cite{kmu10}, which in the Landau gauge is
\begin{equation}
\label{KMHopping}
J_{KM}(\vec{r},\vec{r'})= (-1)^{X+Y+XY} e^{-\frac{\pi}{2}(1-\phi)[X^2+Y^2]} e^{\pi i \phi Y(x+x')},
\end{equation}
with $X=x-x'$ and $Y=y-y'$,
creating the spectrum seen in Fig.\ref{fig:KMButterfly}. This model has a number of remarkable properties. While the hoppings are formally of infinite range, the absolute value of the hopping strength decays as a Gaussian, so unless the flux  $\phi $  is close to $1$ only few nearest neighbors are important. The phase of the hopping strength follows a simple plus/minus pattern apart from the complex coefficients created by the magnetic field. Both of these qualities make the model more conductive to experimental realization \cite{kmu10}.

However, the derivation in Ref.\cite{kmu10} relies on a sum rule, obtained through exploration of the completeness of coherent states for a harmonic oscillator \cite{russianpaper}. Consequently, it is not clear if this Hamiltonian is unique in providing a LLL or wether similar lattice Hamiltonians can be engineered for higher Landau levels or other flat band models. In the next section, we consider the most general model with the same properties and argue that  projection operators clarify the physics of the correspondence between the continuum and the lattice models. 

\section{Projection operators}
\label{sec:Projection}

We start by considering the Hilbert space $\Omega$ formed by all one particle wavefunctions on the lattice, which is spanned by the set $|x,y\rangle$. This Hilbert space can be separated into two disjoint sets; the subspace of wavefunctions $\Omega_0$ that can be written as a linear combination of the discretized LLL wavefunctions $\psi_{k_y}$, and its complement $\Omega'_0$. Then formally we can define the projection operator $P_0$ with
\begin{equation}
P_0 \psi = \left\{ \begin{array}{r@{\quad : \quad}l} \psi & \psi \in \Omega_0, \\ 0 & \psi \in \Omega'_0 . \end{array} \right.
\end{equation}

In terms of this projection operator the following Hamiltonian can be defined
\begin{equation}
\label{MostGeneralForm}
{\cal H} = \epsilon_0 P_0 + \left(1-P_0\right) {\cal H}' \left(1-P_0\right).
\end{equation}
In this definition, $\epsilon_0$ is an arbitrary real number and ${\cal H}'$ is an arbitrary one body Hamiltonian. We see that ${\cal H}$ satisfies the two conditions we set for the long range hopping model with a LLL. All the states in the space $\Omega_0$ have energy $\epsilon_0$, thus they from a massively degenerate manifold. All the wavefunctions obtained by discretization of the LLL wavefunctions lie in $\Omega_0$, thus are part of this degenerate manifold. 
The required long range hopping strengths are obtained by
\begin{equation}
J(\vec{r},\vec{r'})= \langle x,y | {\cal H} | x',y' \rangle.
\end{equation}
The choice of  ${\cal H}'$ affects the real space hoppings without altering the properties of the degenerate manifold. Hence all possible ${\cal H}'$ yield long range hopping models with the same properties of Kapit-Mueller Hamiltonian, showing that there are infinitely many such models.

While the existence of the projector $P_0$ is easily assured, its explicit construction is not straightforward. While two LLL wavefunctions with $k_y \neq k_y'$ are orthogonal,
\begin{equation}
\langle \Psi_{k_y} | \Psi_{k'_y} \rangle = \int d{\cal X} d{\cal Y} \Psi_{k_y}^*\left({\cal X},{\cal Y}\right) \Psi_{k'_y}\left({\cal X},{\cal Y}\right) = 0,
\end{equation}
the functions formed by sampling them on the lattice are in general not so,
\begin{equation}
\langle \psi_{k_y} | \psi_{k'_y} \rangle = \sum_{x,y} \psi_{k_y}^*\left(x,y\right) \psi_{k'_y}\left(x,y\right) \neq 0.
\end{equation}
Thus the projector $P_0$ can not simply be expressed as
\begin{equation}
P_0 \neq \sum_{k_y} | \psi_{k_y} \rangle \langle \psi_{k_y} |.
\end{equation}
Furthermore, as we discuss in section \ref{sec:1LL}, the dimension of the subspace $\Omega_0$ is not necessarily equal to the dimension of the Hilbert space spanned by continuum LLL wavefunctions.

Projection operators defined in this manner are not specific to the LLL, and this method can be easily extended to higher LL, or to mimic continuum Hamiltonians with lattice models in general. However, explicit calculation of the long-range hopping strengths depends on the details of the wavefunctions in the degenerate manifold. In the next section, we construct the projection operator $P_0$ by forming a complete orthonormal set out of the projected LLL wavefunctions and give explicit expressions for hopping strengths. The following section contains the application of the method to the first excited LL, and restrictions involved in discretization.

\section{Lowest Landau Level}
\label{sec:LLL}

The wavefunctions in the LLL are labeled by the wavevector $k_y \in {\mathbb R} $ in the Landau Gauge,
\begin{equation}
\Psi_{k_y}\left({\cal X},{\cal Y}\right) = \left( \frac{1}{\pi \ell^2} \right)^\frac{1}{4} e^{-\frac{({\cal X}-{\cal X}_0)^2}{2 \ell^2}} e^{i k_y {\cal Y}},
\end{equation}
with ${\cal X}_0=\frac{\hbar k_y}{e B}$.  We sample these wavefunctions on the (integer) lattice points $x,y$ to form the discretized wavefunctions
\begin{equation}
\psi_{k_y}\left(x,y\right)= \left(\frac{2 p}{q}\right)^\frac{1}{4} e^{- \frac{\pi p}{q} \left(x-\frac{q}{2 \pi p} k_y\right)^2} e^{i k_y y}.
\end{equation}
Here, the only parameter is the flux quantum per plaquette of the lattice $\phi=\frac{p}{q}$, with $p$ and $q$ co-prime integers. 
After discretization the sampled wavefunctions are no longer normalized, as the inner product changes from an integration to a sum on the lattice. The normalized wavefunctions can be written as
\begin{equation}
\psi_{k_y}\left(x,y\right)= \frac{1}{\sqrt{2 \pi \theta_3\left(i k_y|i \frac{2 p}{q}\right)}} e^{- \frac{\pi p}{q} x^2 + k_y x} e^{i k_y y},
\end{equation}
where $\theta_3\left(z|\tau\right)$ is the third Jacobi Theta function of $z$, with quasiperiod $\tau$ \cite{thetafunctions}. The discretization process also changes the orthogonality properties of the lattice wavefunctions. Continuum wavefunctions within the Landau gauge are orthogonal as  ${\cal Y}$ translation is a continuous symmetry. Upon introduction of the lattice, this continuous symmetry is broken into a discrete translational symmetry, and as a result two lattice wavefunctions $\psi_{k_y}$ and $\psi_{k'_y}$ are no longer orthogonal if their wavevectors differ by an integer multiple of $2 \pi$. Defining a restricted wavevector $k_y=\bar{k}_y+ 2 \pi l_1$ and $k'_y=\bar{k}_y+2 \pi l_2$, with $l_1,l_2$ integers and $-\pi<\bar{k}_y\le \pi$ we calculate the overlap
\begin{widetext}
\begin{equation}
\left\langle \psi_{\bar{k}_y+2 \pi l_1} | \psi_{\bar{k}_y +2 \pi l_2} \right\rangle = e^{- \frac{q}{2 p} \pi (l_1-l_2)^2} \frac{\theta_3\left(\frac{q \bar{k}_y}{2 p}+\frac{\pi q}{2 p} (l_1+l_2) | i \frac{q}{2 p} \right)}{\sqrt{\theta_3\left(\frac{q \bar{k}_y}{2 p}+\frac{\pi q}{p} l_1 | i \frac{q}{2 p} \right) \theta_3\left(\frac{q \bar{k}_y}{2 p}+\frac{\pi q}{p} l_2 | i \frac{q}{2 p} \right)}}.
\end{equation} 
\end{widetext}
As discussed in the previous section, our aim is to form a projection operator $P_0$ into the subspace spanned by all these wavefunctions. The simplest way to achieve this is to orthogonalize the set $\psi$, using the magnetic translation symmetry in the $x$ direction. To illustrate the method with the least possible algebra, we now restrict $p=1$ and assume that $q$ is an even integer. At the end of the section, we discuss the extension to the general case.

For $p=1$ the overlap between two states 
\begin{equation}
\left\langle \psi_{\bar{k}_y+2 \pi l_1} | \psi_{\bar{k}_y +2 \pi l_2} \right\rangle = e^{- \frac{q}{2 } \pi (l_1-l_2)^2 },
\end{equation}
is independent of $\bar{k}_y$. Thus discrete translation symmetry by $q$ steps in the $x$ direction leads to the definition of
\begin{equation}
\label{OrthogonalSet}
\left| \bar{k}_y,k_x \right\rangle = \frac{1}{\sqrt{ 2 \pi \theta_3\left(\frac{k_x}{2}|i\frac{q}{2}\right)}} \sum_{m=-\infty}^{\infty} e^{i k_x m} \left| \psi_{\bar{k}_y+2 \pi m} \right\rangle,
\end{equation}
where $k_x$ is by definition bounded $-\pi < k_x \le \pi$
These states are orthonormal, satisfying
\begin{equation}
\left\langle \bar{k}_y,k_x | \bar{k}'_y,k'_x \right\rangle= \delta(\bar{k}_y-\bar{k}'_y) \delta(k_x-k'_x).
\end{equation}

The projector is then written simply in terms of these states as
\begin{equation}
P_0= \int_{-\pi}^{\pi} dk_x \int_{-\pi}^{\pi} d\bar{k}_y  \left| \bar{k}_y,k_x\right\rangle \left\langle \bar{k}_y,k_x \right|.
\end{equation}
Now we can construct the long range model with the desired properties using $P_0$, the simplest choice is to use the projector $P_0$ as the Hamiltonian
\begin{eqnarray*}
{\cal H} &=& P_0 \\  \nonumber
&=& \sum_{\vec{r},\vec{r}'} J\left(\vec{r},\vec{r}'\right) a^\dagger_{x,y} a_{x',y'},
\end{eqnarray*}
with the hopping strengths given by
\begin{widetext}
\begin{equation}
\label{IntegralHopping}
J(\vec{r},\vec{r}')=-\int_0^{2 \pi} \frac{d k_y}{2 \pi} \int_{-\pi}^{\pi} \frac{d k_x}{2 \pi} \frac{ \theta_3 \left( \frac{k_x+i q k_y}{2} - i \pi x | i q \right) \theta_3 \left( \frac{-k_x+i q k_y}{2} - i \pi x' | i q \right)}{\theta_3 \left( \frac{k_x}{2} | i \frac{q}{2} \right) \theta_3 \left( i k_y | i \frac{2}{q} \right)} e^{-\frac{\pi}{q} \left(x^2+x'^2\right)} e^{k_y \left[(x+x')+i(y-y')\right]}.
\end{equation}
\end{widetext}

This integral is evaluated in the appendix, yielding
\begin{equation}
\label{AnalyticHopping}
J(\vec{r},\vec{r}')= -\frac{q}{4} \frac{e^{i \frac{\pi}{q} (x+x')Y}}{\left[\theta'_1\left(0|\frac{2 i}{q}\right)\right]^2} {\cal M}_{X}\left(Y\right) {\cal M}_{Y}\left(X\right),
\end{equation}
with $X=x-x'$, $Y=y-y'$ and
\begin{equation}
{\cal M}_a(b)= \left \{ \begin{array} {r@{\quad : \quad}l} \frac{\theta_1\left(\frac{\pi b}{q}|\frac{2 i}{q}\right)}{\sinh\left(\frac{\pi b}{2}\right)} & a \quad \mathrm{even} ,\\
\frac{\theta_4\left(\frac{\pi b}{q}|\frac{2 i}{q}\right)}{\cosh\left(\frac{\pi b}{2}\right)} & a \quad \mathrm{odd}. \end{array} \right.
\end{equation}

\begin{figure}[ht]
\includegraphics[scale=0.6]{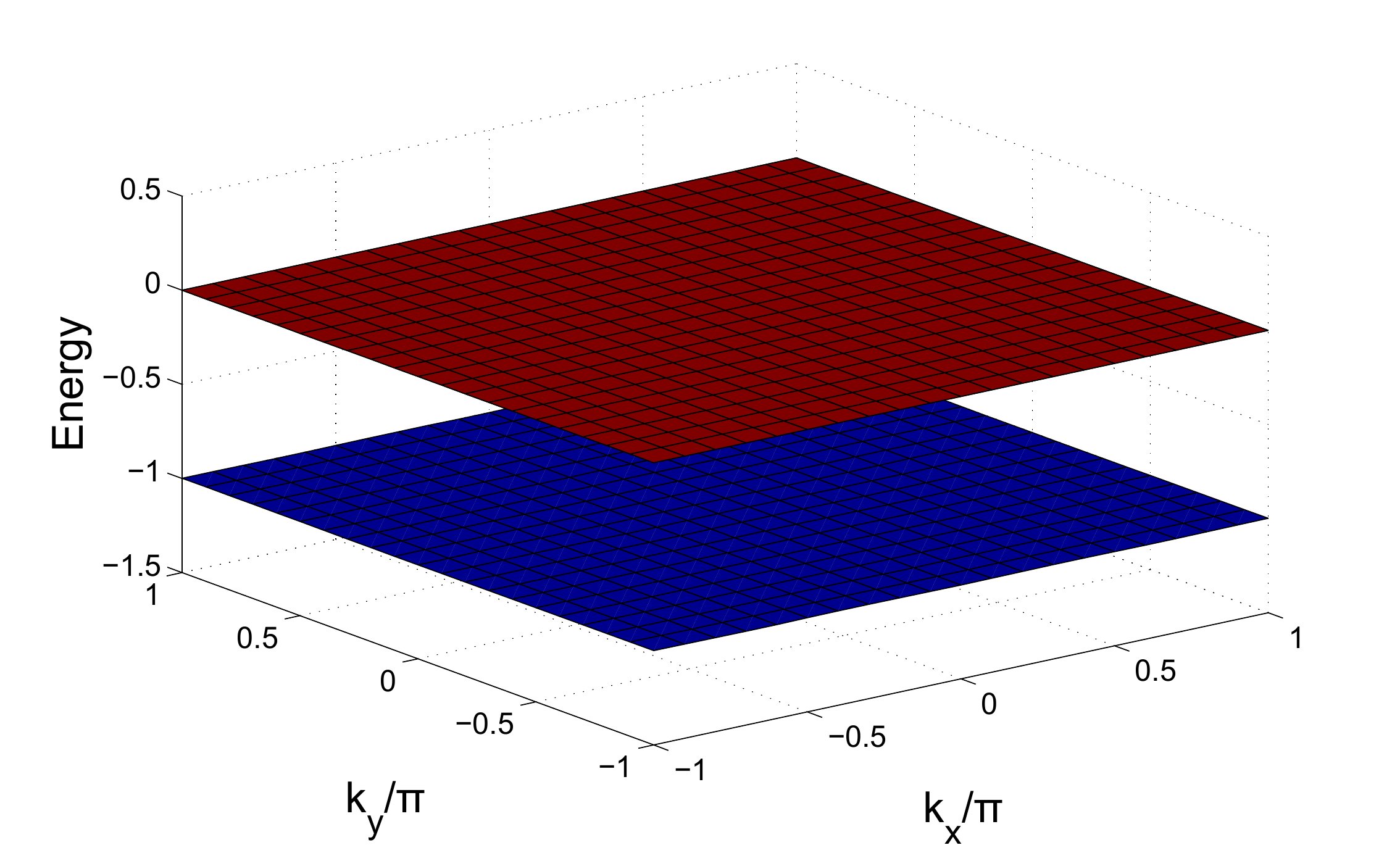}
\caption{\label{fig:flatbands} (Color online) Energy bands as a function of  $k_x,k_y$, for $\phi=1/3$,  on the square lattice with long range hopping given in Eq.\ref{AnalyticHopping}. The lower band is the LLL manifold on the lattice, while the upper band which contains twice as many states is formed by all the remaining states. Both bands are flat in $k$ space to our numerical precision, verifying that the constructed Hamiltonian is the desired projection operator $P_0$. The flatness of the bands should be compared with Fig. \ref{fig:curvedbands}}.
\end{figure}

\begin{figure}[ht]
\includegraphics[scale=0.6]{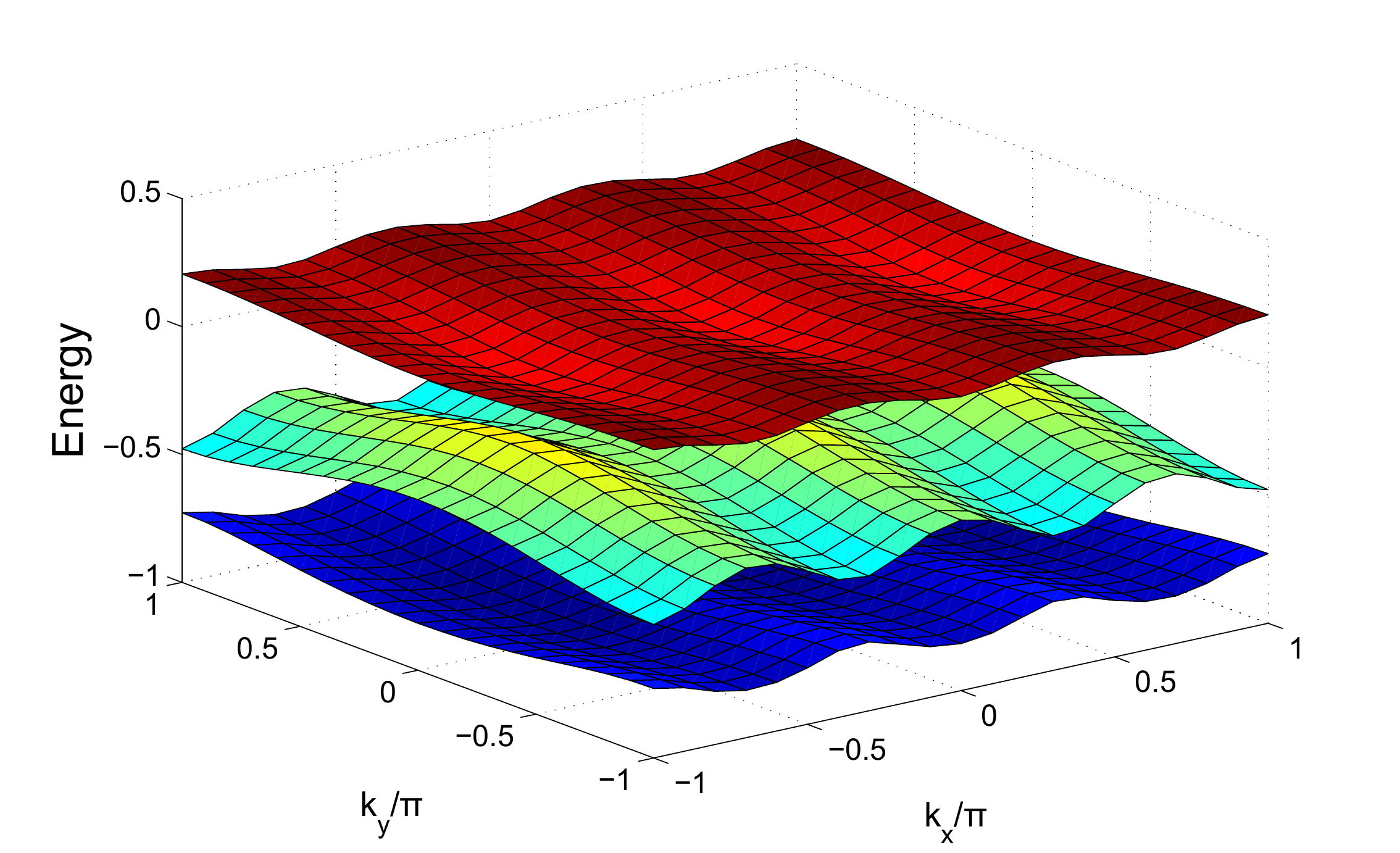}
\caption{\label{fig:curvedbands} (Color online) Energy as a function of $k_x,k_y$, for a square lattice with nearest hopping at $\phi=1/3$. Notice that the width of each band is comparable to the separation between the bands unlike Fig.\ref{fig:flatbands} }
\end{figure}

We calculated the energy spectrum for a lattice with these hopping strengths, for different values of $\phi=1/q$. We see that the lowest band, with energy $-1$, is flat to our numerical precision. The bands for our hopping strengths (Fig \ref{fig:flatbands}) should be contrasted with the usual nearest neighbor hopping bands for $\phi=1/3$ (Fig. \ref{fig:curvedbands}). We performed another numerical check to make sure that the lattice LLL wavefunctions are eigenfunctions of this potential by directly applying the Hamiltonian to a few LLL wavefunctions and found that the deviation of their energy from $-1$ is less than our numerical precision ($10^{-7}$ in our dimensionless units).

We also explored how the Hofstadter butterfly spectrum evolves to the flat band spectrum using the following method. We defined a set of hoppings that interpolates from the nearest neighbor hopping to our long range hopping model
\begin{equation} 
\label{CollapseHopping}  
\tilde{J}_\alpha\left(\vec{r},\vec{r}'\right) =  \left \{ \begin{array} {r@{\quad : \quad}l} \alpha J\left(\vec{r},\vec{r}'\right) & |\vec{r}-\vec{r}'|>1 \\
J\left(\vec{r},\vec{r}'\right) & |\vec{r}-\vec{r}'| \le 1 \end{array}  \right. ,
\end{equation}
for $0 \le \alpha \le 1$. The result is displayed in figure 4 for $\phi=1/3$, clearly showing the collapse of the lowest Hofstadter band to the LLL. The two higher bands  coalesce into another flat band, as expected from the projector nature of our Hamiltonian.

\begin{figure}[ht]
\includegraphics[scale=0.55]{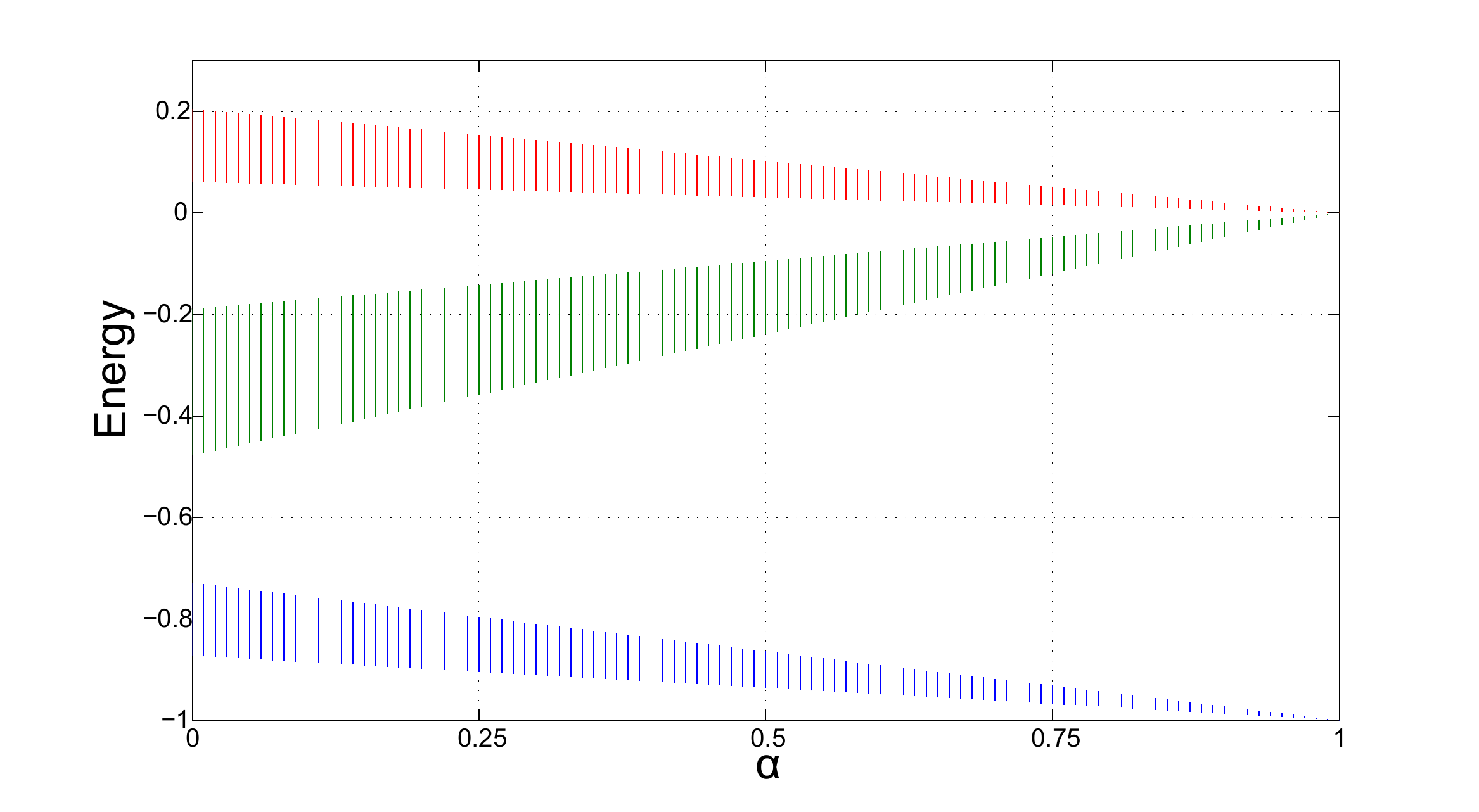}
\caption{\label{fig:Collapsep1q3} (Color online) Energy bands of the square lattice at $\phi=1/3$, as a function of dimensionless $\alpha$ defined in Eq.\ref{CollapseHopping}. The lowest band of the Hofstadter model (at $\alpha=0$) narrows as long range hopping is turned on and collapses to the LLL at $\alpha=1$. As the lowest band does not touch the other bands during evolution, the first Chern number of the band is preserved. This Chern number is equal to $1$, the Hall conductivity for the LLL. Notice that the two higher bands also shrink down to a point as the Hamiltonian defined by Eq.\ref{AnalyticHopping} projects out the states in these bands. The energies at $\alpha=0$ are shifted from the Hofstadter model by adding a constant on site energy term $J(0,0)=1/3$ so that the LLL has energy equal to $-1$.}
\end{figure}

\begin{figure}[ht]
\includegraphics[scale=0.55]{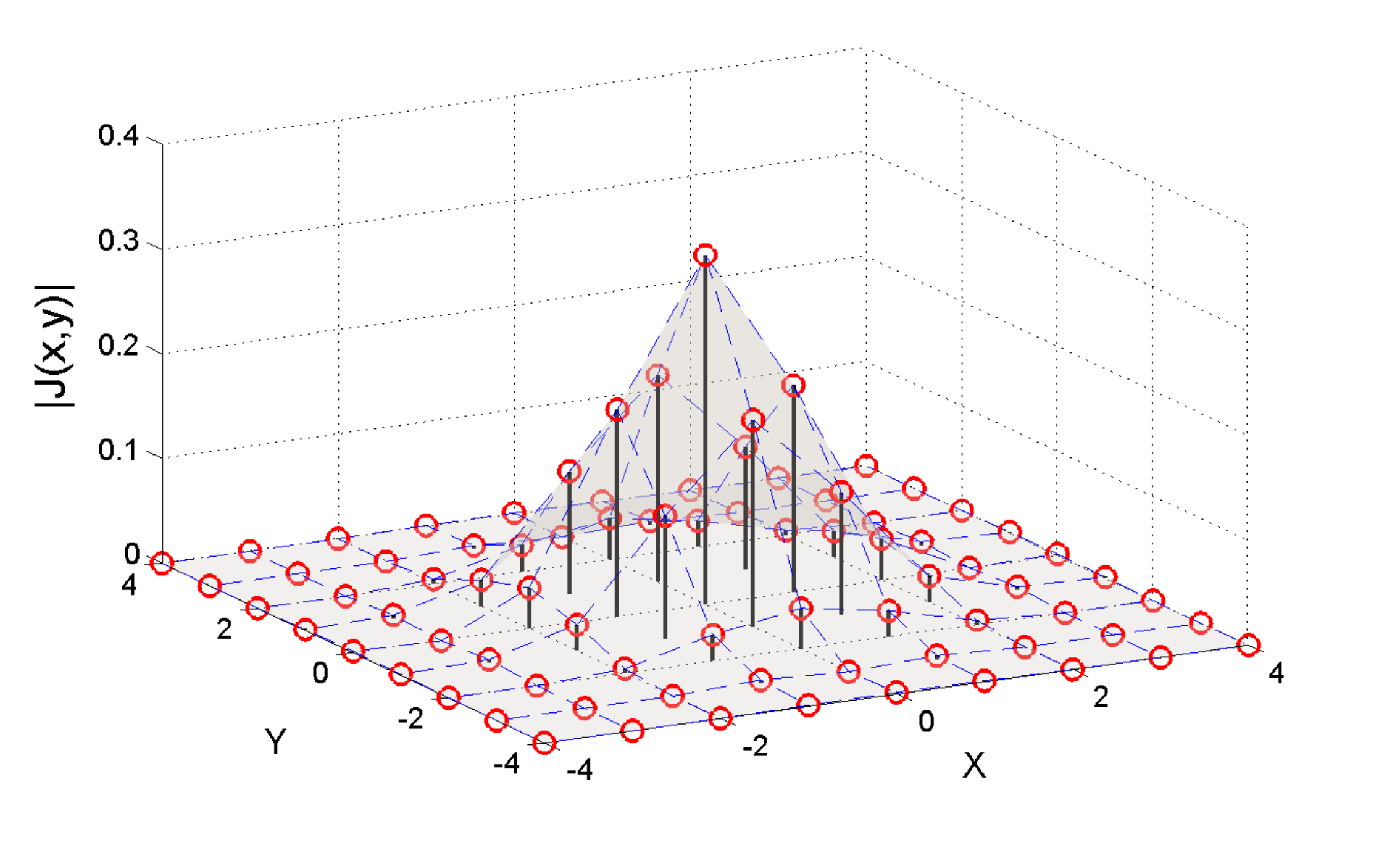}
\caption{\label{fig:LLLHopping} (Color online) Absolute value of hopping strengths from the origin to the nearest sites, $|J(x,y)|$, defined in Eq.\ref{AnalyticHopping}. The hopping strengths decay exponentially for all $\phi$, while for the case of $\phi=1/3$ shown above they are negligible beyond the third nearest neighbor. }
\end{figure}

An important property of the long range hopping model presented above is the fast decay of the hopping parameters. The hopping strengths decay exponentially with distance, and for all $\phi=1/q$, even the third nearest neighbor hopping is almost negligible.  In figure \ref{fig:LLLHopping}, we plot the absolute value of the hopping strengths. The finite range of the hopping increases the possibility of experimental realization of lattices with a LLL, and thus cold atom fractional quantum Hall systems. One can also further optimize the most general Hamiltonian Eq.\ref{MostGeneralForm} so that LLL is achieved with the introduction of hopping to minimum number of sites.

It is also worth noting that all the orthogonal states found in Eq.\ref{OrthogonalSet} are normalizable. Because such an orthogonal basis can be labeled by $k_y$, it is proved that all the discretized LLL wavefunctions are linearly independent. So for the LLL, for all values of $\phi=1/q$ the dimension of the Hilbert space spanned by the projected wavefunctions is equal to the dimension of the Hilbert space spanned by continuum LLL wavefunctions.

Lifting the restriction $p=1$, the same orthogonalization method can be applied. However now the discrete translational symmetry for sampled LLL states is reduced p-fold. Hence, one must modify the definition of the states in the orthonormal basis spanning the lattice LLL by introducing a band index,
\begin{equation}
\left| \bar{k}_y,k_x,\gamma \right\rangle = \sum_{r=-\infty}^{\infty} \sum_{s=0}^{p-1} e^{i k_x r} u_\gamma(s) \left| \psi_{\bar{k}_y+2 \pi (m p+s)} \right\rangle,
\end{equation}
with $\gamma$ defining the corresponding magnetic band. The eigenvector $u_\gamma(s)$ can be obtained by diagonalizing the $p$ by $p$ matrix 
\begin{widetext}
\begin{equation}
\label{AMatrix}
A_{s_1,s_2}=e^{-\frac{q \pi}{2 p} (s_1-s_2)^2} \frac{\theta_3\left( \frac{q \bar{k_y}}{2 p} + \frac{q \pi}{2 p} (s_1+s_2) | \frac{i q}{2 p} \right)
 \theta_3\left( \frac{k_x}{2} - i \frac{q \pi}{2 } (s_1-s_2) | \frac{i q p}{2} \right)}{\theta_3\left( \frac{q \bar{k_y}}{2 p} + \frac{q \pi}{p} s_1 | \frac{i q}{2 p} \right) \theta_3\left( \frac{q \bar{k_y}}{2 p} + \frac{q \pi}{p} s_2 | \frac{i q}{2 p} \right)}.
\end{equation}
\end{widetext}
While it's possible to express the long range hopping strengths in terms of the inverse of the matrix $A$, we have not been able to analytically evaluate this inverse for a general value of $p$. Even for $p=2$ the resulting expressions are very long and provide little insight. It should be noted  that it's numerically very easy to calculate the hopping strength for any lattice separation. Again numerical diagonalization of the matrix $A$ for arbitrary values of $k_x,k_y$ lead us to conclude that as long as $\phi<1$ the lattice LLL Hilbert space dimension is equal to the continuum LLL Hilbert space dimension. At exactly $\phi=1$ the number of LLL states is equal to the number of lattice sites, and the space spanned by them is the total Hilbert space of the lattice. 

We also calculated the energy spectrum of a lattice with the hoppings given in Eq.\ref{AnalyticHopping} for general values of $\phi$ by replacing $q \rightarrow 1/\phi$. In the resulting energy spectrum, we obtain two distinct bands, and while the LLL band is not degenerate, kinetic energy is quenched to such a degree that the width of the band is always smaller than $3 \%$ of the separation between the bands. See figure 6.

\begin{figure}[ht]
\includegraphics[scale=0.6]{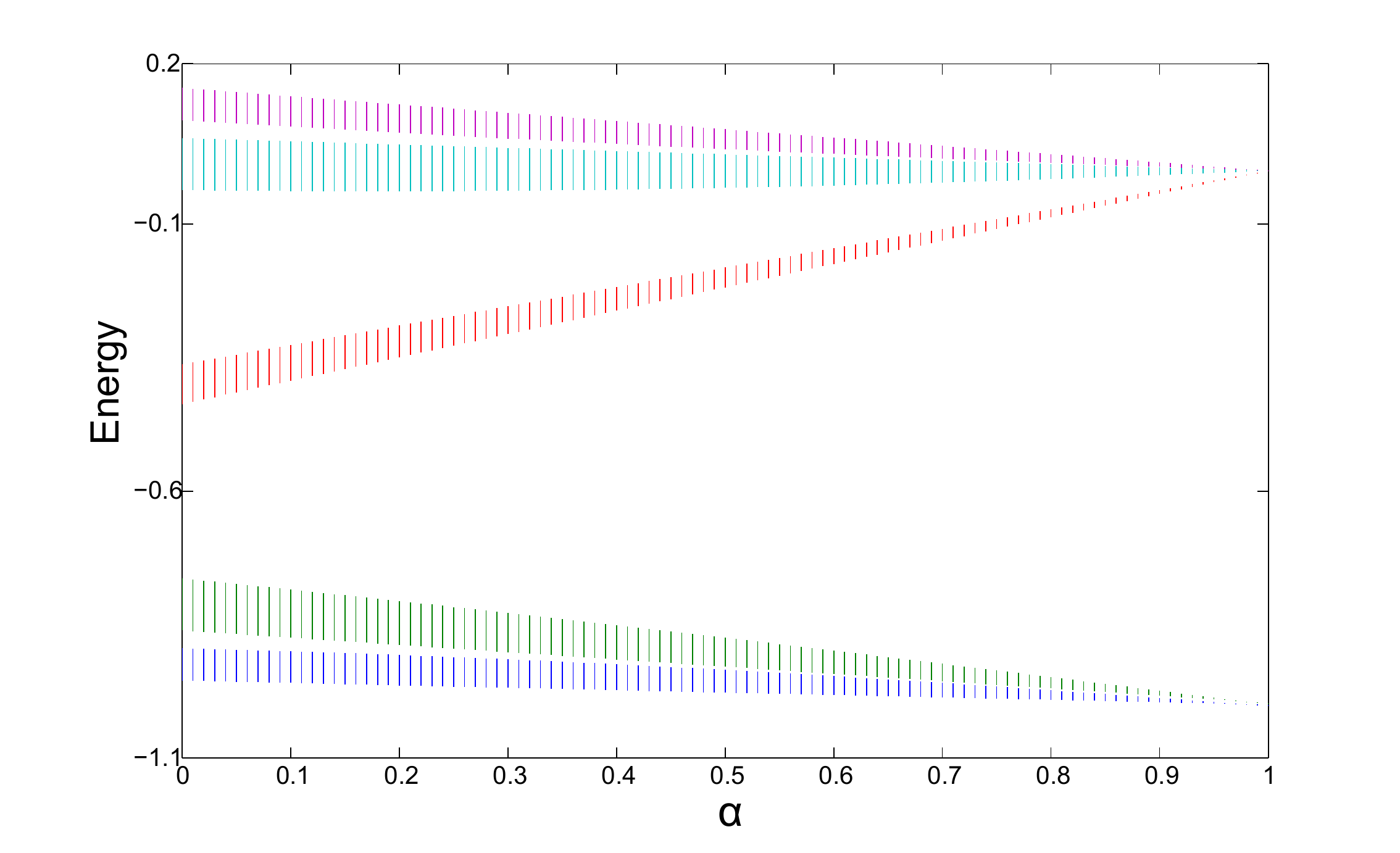}
\caption{\label{fig:Collapsep2q5} (Color online) Collapse of the bands into narrow manifolds for $\phi=2/5$ as a function of $\alpha$ (See Eq.\ref{CollapseHopping}). Here we used the analytic expressions obtained for $\phi=1/q$ by replacing $q\rightarrow 1/q$, thus the collapse of the bands into the LL structure is not exact, nonetheless the final width of the manifolds are below $10^{-3}$ in our dimensionless units. Exactly degenerate LLL can be obtained through by inverting the matrix in Eq.\ref{AMatrix}, and numerically calculating the resulting integral. }
\end{figure}

\section{ Higher Landau Levels} 
\label{sec:1LL}

The method presented in the previous chapter does not depend on the properties of the LLL, and can easily be extended to higher LL. As an example, we consider the first excited LL (1LL). In the Landau gauge, the wavefuncions of the 1LL are
\begin{equation}
\Psi_{1,k_y}\left({\cal X},{\cal Y}\right)= \frac{1}{\left( 4 \pi \ell^2 \right)^\frac{1}{4}} 2 \left(\frac{{\cal X}-\frac{\hbar k_y}{e B}}{\ell}\right) e^{-\frac{1}{2} \left(\frac{{\cal X}-\frac{\hbar k_y}{e B}}{\ell}\right)^2} e^{i k_y {\cal Y}}.
\end{equation}
Upon sampling on the square lattice, we obtain
\begin{equation}
\psi_{1,k_y}\left(x,y\right) = \frac{\left({\cal A}_0(k_y)\right)^{-\frac{1}{2}}}{\sqrt {2\pi}} \left(x - \frac{q k_y}{2 \pi} \right) e^{-\frac{\pi}{q}\left(x-\frac{q k_y}{2 \pi}\right)^2} e^{i k_y y},
\end{equation}
where the normalization factor is defined as
\begin{widetext}
\begin{equation}
{\cal A}_r\left(k_y\right)=\sum_{x=-\infty}^{\infty} \left(x-\frac{q k_y}{2 \pi}\right) \left(x-\frac{q k_y}{2 \pi} + q r \right) e^{-\frac{\pi}{q} \left[\left(x-\frac{q k_y}{2 \pi}\right)^2 + \left(x-\frac{q k_y}{2 \pi} + q r \right)^2 \right] }.
\end{equation}
The normalization factors can be expressed in terms of the Jacobi theta functions and their derivatives as
\begin{equation}
{\cal A}_0\left(k_y\right)=e^{-\frac{q k^2_y}{2\pi}} \left[-\frac{1}{4} \theta''_3\left(i k_y|\frac{2i}{q}\right)+\frac{i q k_y}{2\pi} \theta'_3\left(i k_y|\frac{2i}{q}\right)+\frac{q^2 k_y^2}{4\pi^2} \theta_3\left(i k_y|\frac{2i}{q}\right)\right]
\end{equation}
\end{widetext}

\begin{figure}[ht]
\includegraphics[scale=0.55]{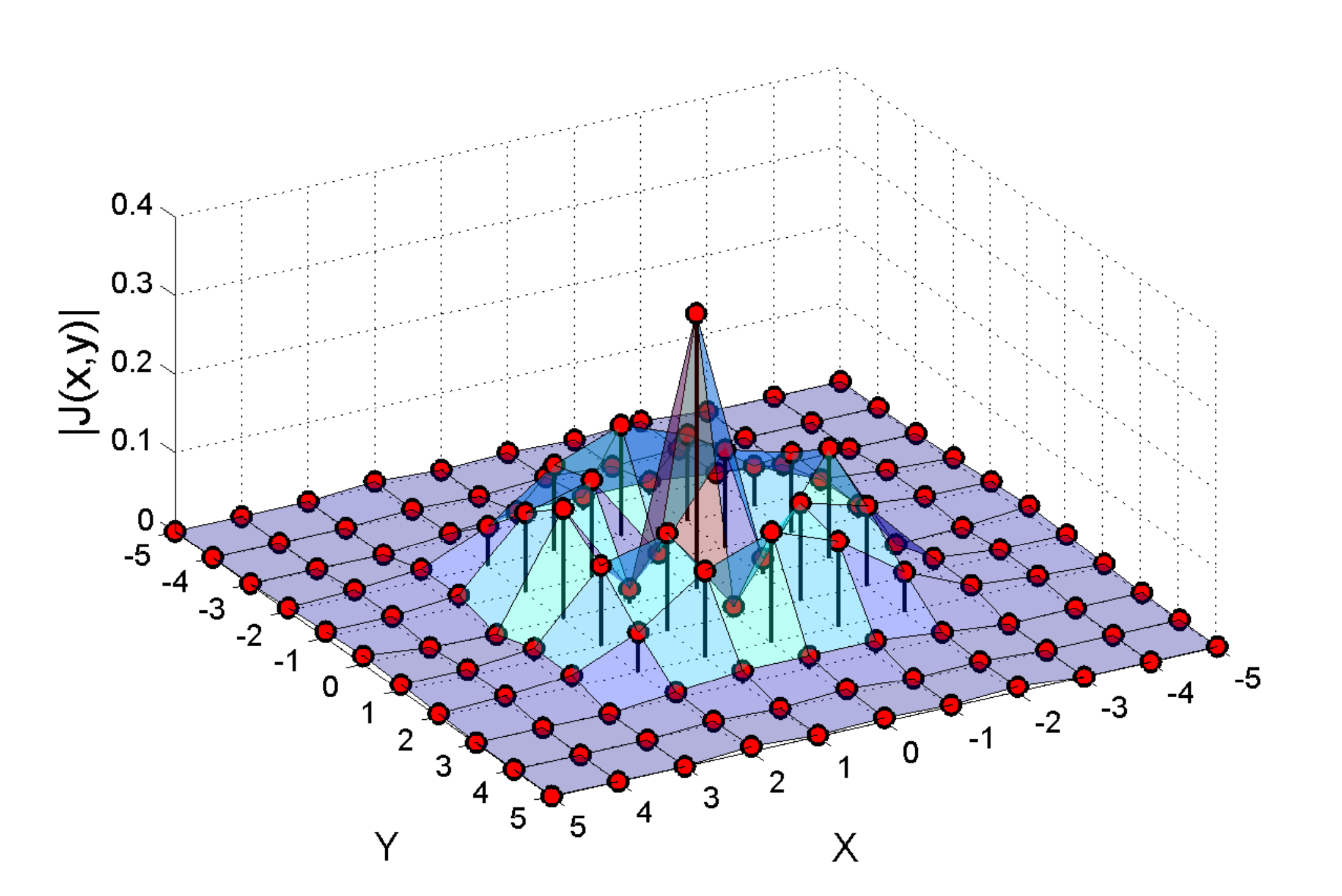}
\caption{\label{fig:1LLHopping} (Color online) Absolute value of the hopping strengths that reconstruct the 1LL at $\phi=1/3$. Notice that the hopping strengths decay slower than the hoppings needed for the LLL construction. It is also remarkable that the overlaps for the next nearest neighbors is larger in magnitude compared to the nearest neighbors. The hopping strengths are calculated by numerical evaluation of Eq.\ref{1LLHopping}}
\end{figure}

To find the long range hopping model that has a massively degenerate first LL, we  construct the operator $P_1$, which projects into the Hilbert space spanned by the all the $|\psi_{1,k_y}\rangle$. This can again be achieved by forming an orthogonal set from linear combinations of these sampled states based on the discrete translational invariance. For mathematical simplicity we again take $p=1$, in which case the overlap between two $k_y$ states differing by an integer multiple of $2 \pi$ is
\begin{equation}
\left\langle \psi_{\bar{k}_y+2 \pi l_1}| \psi_{\bar{k}_y+ 2 \pi l_2} \right\rangle=\frac{1}{2\pi} \frac{{\cal A}_{l_1-l_2}(\bar{k}_y)}{{\cal A}_0(\bar{k}_y)}.
\end{equation}
The orthogonal set can be formed by straightforward Fourier transformation
\begin{equation}
\label{1LLOrthagonal}
|\bar{k}_y,k_x\rangle = \sum_{l_x} e^{i k_x l_x} |\psi_{\bar{k}_y+2 \pi l_x} \rangle,
\end{equation}
yielding
\begin{equation}
\langle \bar{k}'_y,k'_x | \bar{k}_y,k_x \rangle = \delta\left(\bar{k}'_y - \bar{k}_y\right) \delta(k'_x-k_x) \sum_{r} e^{i k_x r} \frac{{\cal A}_r(\bar{k}_y)}{{\cal A}_0(\bar{k}_y)}
\end{equation}
The projector $P_1$, and hence the long range hopping Hamiltonian can be constructed out of these wavefunctions 
\begin{equation}
P_1= \int_{-\pi}^{\pi} \int_{-\pi}^{\pi} d\bar{k}_y dk_x |\bar{k}_y,k_x\rangle\langle\bar{k}_y,k_x|,
\end{equation}
which in the lattice basis translates into the hopping strengths
\begin{widetext}
\begin{equation}
\label{1LLHopping}
J_1(\vec{r},\vec{r'})=\int_0^{2\pi} d k_y \int_{-\pi}^{\pi} d k_x \sum_{\ell_1,\ell_2=-\infty}^{\infty}  \frac{e^{i k_x (\ell_1-\ell_2)}}{2\pi {\cal S}_{(k_x,k_y)}} \left(x-\frac{q k_y}{2\pi}-q \ell_1\right) \left(x'-\frac{q k_y}{2\pi}-q \ell_2\right) e^{i k_y (y-y')-\frac{\pi}{q}[(x-\frac{q k_y}{2\pi}-q \ell_1)^2+(x'-\frac{q k_y}{2\pi}-q \ell_2)^2]},
\end{equation}
\end{widetext}
where ${\cal S}=\sum_{r=-\infty}^{\infty} e^{-i k_x r}{\cal A}_r(k_y). $
We numerically verified that discretized 1LL wavefunctions are eigenfunctions of the Hamiltonian with these hopping strengths, forming a degenerate band as intended. Furthermore as displayed in figure \ref{fig:1LLHopping}, these hopping strengths also fall of quickly beyond the first few neighbors, similar to the LLL. We remark that the most general form Eq \ref{MostGeneralForm} can serve as a starting point to find hopping strengths with even shorter ranges.

\begin{figure}[ht]
\includegraphics[scale=0.55]{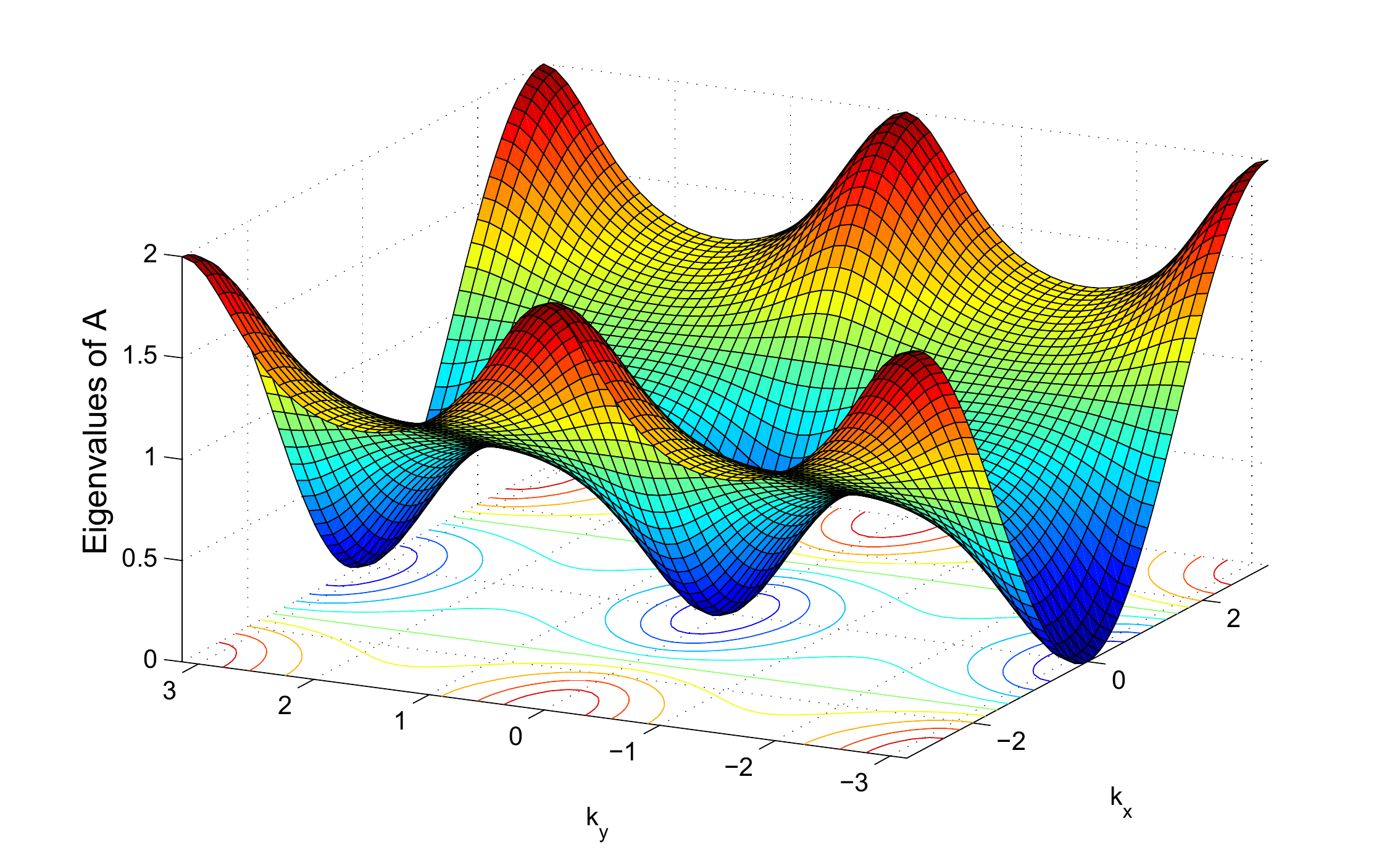}
\caption{\label{fig:OverlapMatrix} (Color online) Eigenvalues of the overlap matrix for 1LL at $\phi=1/2$. Notice that the eigenvalues become zero at two points in the Brillouin zone, $k_x=0,k_y=0$ and $k_x=0,k_y=\pi$. Beyond $\phi>1/2$ there are always eigenvalues equal to zero showing that the set of discretized 1LL are not linearly independent.}
\end{figure}

There is one important distinction between the lattice versions of the LLL and 1LL created by the long range hopping models we give. For the LLL, the discretized set $|\psi_{k_y}\rangle$ form a linearly independent set as long as the flux per plaquette is below one $\phi <1 $. In this sense, the dynamics of the continuum LLL is mimicked exactly by these states, even in the presence of local interactions. This is not the case for 1LL, the set $|\psi_{1,k_y}\rangle$ is linearly independent only if $\phi<\frac{1}{2}$, while for larger flux the number of independent states in the lattice 1LL is smaller than the dimension of the continuum 1LL. This can be deduced by investigating the orthogonal set of states we constructed (see Eq.\ref{1LLOrthagonal}). These states diagonalize the matrix formed by  the overlaps $\langle \psi_{1,k_y'} | \psi_{1,k_y} \rangle $. Thus the eigenvalues of this matrix are
\begin{equation}
{\cal O}(\bar{k}_y,k_x)=\sum_{r} \frac{{\cal A}_r(\bar{k}_y)}{{\cal A}_0(\bar{k}_y)} e^{i k_x r},
\end{equation}
which is plotted in figure \ref{fig:OverlapMatrix} for $\phi=1/2$. At this flux, some of the eigenvalues of the overlap matrix becomes zero, showing that the set of states is no longer linearly independent.

It is easy to see why the lattice sampled wavefunctions do not have the full dimension of the continuum LLs. Flux quantum per plaquette of the lattice $\phi$, is also a measure of the number of states in a given LL per lattice site. For example, at exactly $\phi=1$, the number of states in the LLL is equal to the number of lattice sites. Thus, for any $\phi>\frac{1}{2}$ there are not enough states in the lattice to fully represent both the LLL and the 1LL. At  $\phi=1/2$ all the states in the lattice are either in the space spanned by discretized LLL or discretized 1LL, correspondingly at this flux $P_0+P_1 =1$. It is worth noticing the parallel between our method of lattice discretization and the Nyquist theorem which states that any function of time can be reconstructed if its sampled with a frequency above twice the cutoff frequency of the Fourier transformation of that function \cite{nyquist}. 

Through this reasoning, we can argue that lattice discretized wavefunctions of the $n^{th}$ LL can form a linearly independent set only up to $\phi=\frac{1}{n+1}$. We numerically checked this for the first two LL and observed that this upper limit is reached. While we have not proved that the states in higher LL stay linearly independent up to the limiting $\phi$ value, investigation of the Hofstadter butterfly spectrum and the first Chern numbers related with the gaps strongly suggest this is the case.

The first Chern number for any given band can be defined as the number of times the phase of the wavefunction at an arbitrary point in real space winds around $2 \pi$ as the first Brillouin zone is encircled in reciprocal space \cite{TKKN}. As this arbitrary point in real space may be chosen to lie on the lattice points, our discretization method conserves the Chern number of the  bands. All the LLs in continuum have their dimensionless Hall conductivity equal to $1$, thus we expect the discretized bands to have the same value for their first Chern number. For the LLL, the continuous transformation between the nearest neighbor (Hofstadter) model and our long range hopping model, as displayed in Fig. \ref{fig:Collapsep1q3}, shows that the first Chern number is conserved as the lowest Hofstadter band for $\phi=1/q$ collapses into the discrete LLL. More generally, for $\phi=\frac{p}{q}$ it is the lowest $p$ bands that merge into the LLL.

\begin{figure}[ht]
\includegraphics[scale=0.6]{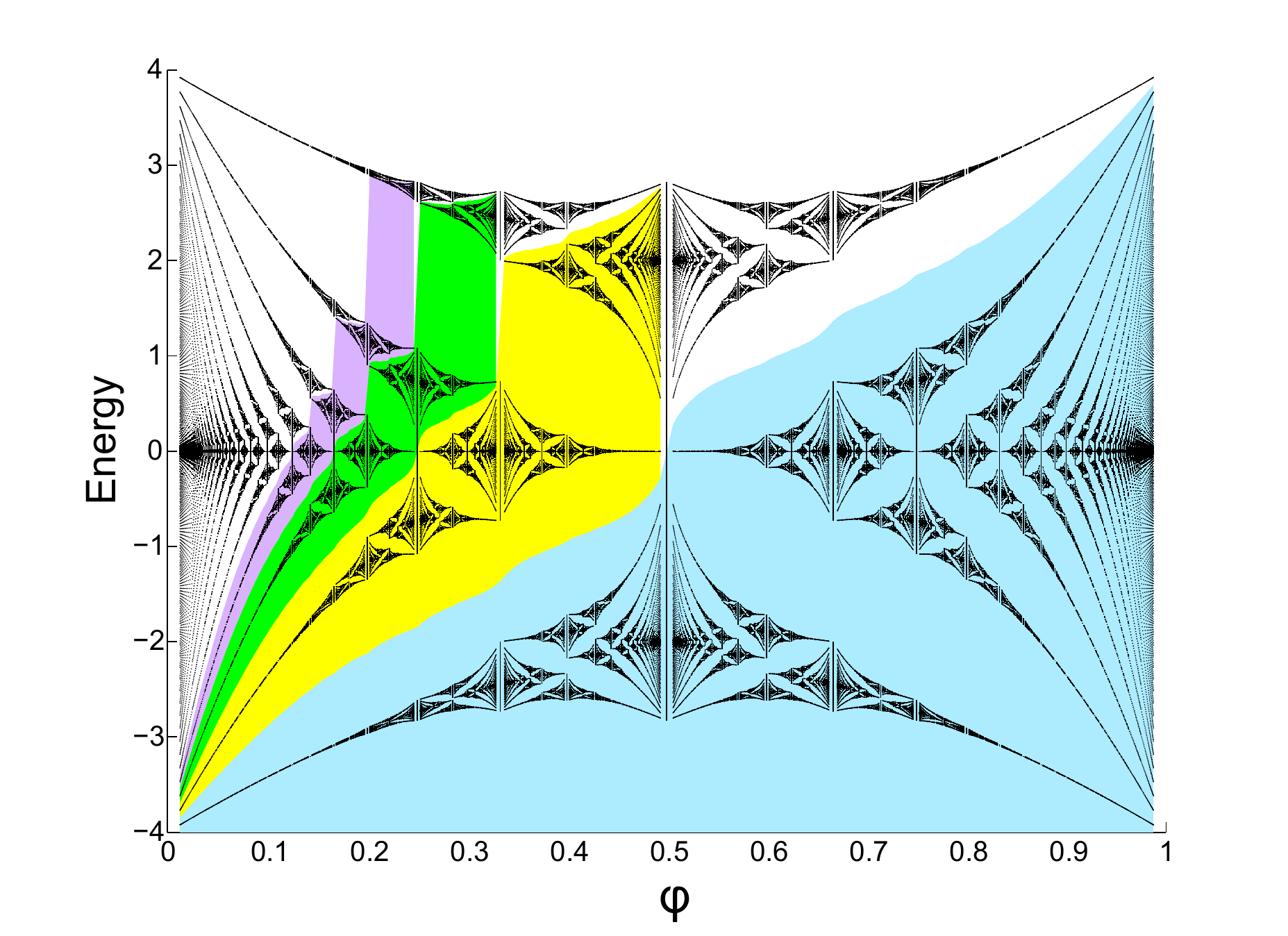}
\caption{\label{fig:LandauLevelButterfly} (Color online) Highlighted regions contain the bands that are completely within the manifold spanned by wavefunctions in a particular LL. The first four LL are shown, LLL (blue), first excited LL (yellow), second excited LL (green), third excited LL (purple). For the $n^{th}$ LL, the region for which discretized LL states are linearly independent is $\phi<\frac{1}{n+1}$.}
\end{figure}

The Hofstadter Butterfly spectrum has clearly identifiable LLs only for very low $\phi \ll 1$, but at a general value of the flux the correspondence between the LLs and the magnetic bands of the spectrum is not clear. A mathematical correspondence can be made by asking the following question: ``Which bands of the Hofstadter butterfly consist entirely of states that lie in the Hilbert space spanned by discretized LL wavefunctions at the same flux?".  While this correspondence is not a guarantee that the physical properties of such bands will be directly inherited from the corresponding LL, it provides an unambiguous way to define the LL in the Hofstadter butterfly.

The Chern numbers, and correspondingly the Hall conductivities of the gaps in the Hofstadter Butterfly are easily determined through a Diophantine equation \cite{TKKN}.  We numerically find that these Chern numbers make it possible to identify the correspondence between the bands and the LLs. While we only could check this extensively for the first two Landau levels, numerical evidence suggests that for $\phi<\frac{1}{n+1}$, the bands lying between the largest gaps with Hall conductivity $n$ and $n+1$ are entirely within the Hilbert space spanned by discretized $n^{th}$ LL wavefunctions. In figure \ref{fig:LandauLevelButterfly}, we mark the regions that are associated with the first four LL on the Hofstadter butterfly. For $\phi>\frac{1}{n+1}$ the Hilbert space spanned by the discretized wavefunctions is smaller than the corresponding LL, which makes it impossible to use the Chern numbers and our long range model construction to determine if the remaining bands in the Hofstadter butterfly can be classified as belonging to a LL. 

\section{Conclusion}
\label{sec:Conclusion}

Simulation of fractional quantum Hall effect and other correlated states using cold atoms remains an important goal. In this paper, we investigated how lattices with long range hopping can be designed to aid that goal by quenching lattice kinetic energy of atoms. Our main results can be summarized as follows.  

We find that there are infinitely many choices for hopping strengths that give a massively degenerate manifold formed out of the lattice sampled LLL wavefunctions. All of these can be constructed starting from the projection operator for this manifold. The hopping strengths for the projector given in Eq. \ref{AnalyticHopping}, decay very quickly with distance similar to the model of Ref.\cite{kmu10}. This fast decay, and recent ideas on modification of tunneling parameters in optical lattices leads us to believe that experimental realization of an optical lattice with a LLL is within reach. An important problem would be to find hopping parameters so that the LLL can be constructed by methods that are currently used in optical lattices by coupling the minimum number of neighbors. 

We also show that the projection operator approach can be applied to higher Landau levels, and find that the hoppings that create a degenerate 1LL manifold also decay exponentially in space. We find that the one to one correspondence between the continuum LL and discretized wavefunctions of the $n^{th}$ LL last only up to a flux of $\phi=\frac{1}{n+1}$. We suggest a method to unambiguously identify LL in the nearest neighbor Hofstadter model, based on a continuous mapping between this model and the calculated long range hoppings. Furthermore we found numerical evidence that this identification is easily accomplished by investigating the first Chern numbers of the gaps in the Hofstadter butterfly.

Our main point in this paper is that introduction of long range hopping to optical lattices will increase the possibility of observation of fractional quantum Hall effects in cold atom systems. It is, however important to notice that the method introduced in this paper, as well as the constructed projection operators have the potential to be used as tools for numerical simulation of correlated states. In a recent paper, braiding statistics of lattice bosons were numerically demonstrated using the long range hopping model.  Instead of the usual exact diagonalization calculations which rely on conserved quantities to limit the size of the Hilbert space, projection to the relevant lattice states can be achieved using the long range hopping model \cite{kgm12}. It would be interesting to extend this method to lattices with non-abelian gauge fields \cite{goldman}.

\acknowledgments
MOO is supported by TUBITAK Grant No. 112T974. MOO wishes to thank the Aspen Center for Physics, the Simons Foundation and the NSF Grant No. 1066293 for hospitality during his visit.  HA wishes to thank Bilkent University for undergraduate comprehensive scholarship.

\appendix
\section{Evaluation of the integral}

Jacobi theta functions of argument $z$ and quasiperiod $\tau$ are defined  following \cite{thetafunctions} as:
\begin{eqnarray}
\theta_1(z|\tau) &=& \sum_{n=-\infty}^\infty (-1)^{(n-1/2)} e^{i\pi\tau(n+1/2)^2} e^{(2n+1)i z}, \nonumber \\
\theta_2(z|\tau) &=& \sum_{n=-\infty}^\infty e^{i\pi\tau(n+1/2)^2} e^{(2n+1)i z}, \nonumber \\
\theta_3(z|\tau) &=& \sum_{n=-\infty}^\infty e^{i\pi\tau n^2} e^{2n i z}, \nonumber \\ 
\theta_4(z|\tau) &=& \sum_{n=-\infty}^\infty (-1)^{n} e^{i\pi\tau n^2} e^{2n i z}.
\end{eqnarray}
  
The integral for the hopping strengths Eq.\ref{IntegralHopping} can be transformed using the identity
\begin{eqnarray}
\theta_3(u|\tau)\theta_3(v|\tau) &=& \theta_3(u+v|2\tau)\theta_3(u-v|2\tau) \nonumber \\
& &+\theta_2(u+v|2\tau)\theta_2(u-v|2\tau),
\end{eqnarray}
to obtain
\begin{equation}
J(\vec{r},\vec{r}')=e^{-\frac{\pi}{q} \left(x^2+x'^2\right)} \left[I_5 I_1+I_6 I_2 \right],
\end{equation} 
where $I_5, I_1, I_6$ and  $I_2$ are the integrals with only one variable (either $k_x$ or $k_y$) given by
\begin{eqnarray}
I_5&=&\int_0^{2 \pi} \frac{d k_y}{2 \pi} \frac{e^{k_y \left[(x+x')+i(y-y')\right]}\theta_3 \left(i q k_y-i \pi (x+x')|2i q\right)}{\theta_3\left(i k_y|\frac{2 i}{q}\right)}, \nonumber \\
I_1&=&\int_{-\pi}^{\pi} \frac{d k_x}{2 \pi} \frac{\theta_3\left(k_x-i \pi (x-x')|2 i q\right)}{\theta_3 \left(\frac{k_x}{2}|i \frac{q}{2}\right)},
\nonumber \\
I_6&=&\int_0^{2 \pi} \frac{d k_y}{2 \pi} \frac{e^{k_y \left[(x+x')+i(y-y')\right]}\theta_2 \left(i q k_y-i \pi (x+x')|2i q\right)}{\theta_3\left(i k_y|\frac{2 i}{q}\right)}, \nonumber \\
I_2&=&\int_{-\pi}^{\pi} \frac{d k_x}{2 \pi} \frac{\theta_2\left(k_x-i \pi (x-x')|2 i q\right)}{\theta_3 \left(\frac{k_x}{2}|i \frac{q}{2}\right)}.
\end{eqnarray}

We use the identity 
\begin{equation}
\theta_2(2z|4\tau)=\frac{1}{2} \left[\theta_3(z|\tau)-\theta_4(z|\tau)\right],
\end{equation} to write
\begin{eqnarray}
I_1&=&I_3+I_4, \\ \nonumber
I_2&=&I_3-I_4,
\end{eqnarray}
where $I_3$ and $I_4$ are given by  
\begin{eqnarray}
I_3&=&\int_{-\pi}^{\pi} \frac{d k_x}{4 \pi} \frac{\theta_3\left(\frac{k_x}{2}-i \frac{\pi (x-x')}{2}| i \frac{q}{2}\right)}{\theta_3 \left(\frac{k_x}{2}|i \frac{q}{2}\right)}, \\
I_4&=&\int_{-\pi}^{\pi} \frac{d k_x}{4 \pi} \frac{\theta_4\left(\frac{k_x}{2}-i \frac{\pi (x-x')}{2}| i \frac{q}{2}\right)}{\theta_3 \left(\frac{k_x}{2}|i \frac{q}{2}\right)}.
\end{eqnarray}

\begin{figure}[ht]
\includegraphics[scale=0.6]{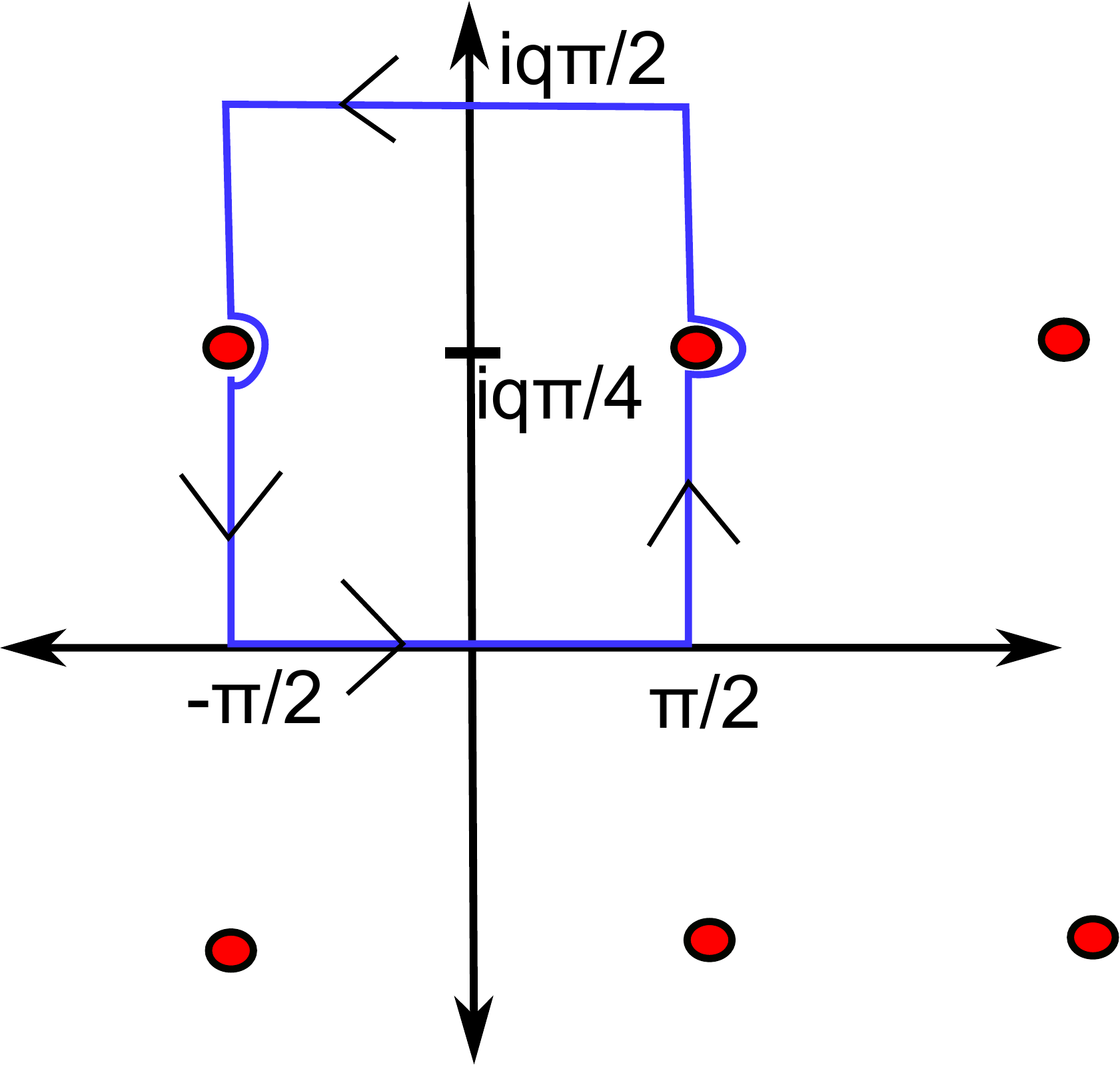}
\caption{\label{fig:I3Contour} (Color online) Contour for the integral $I_3$ is shown in the complex $k_x$ plane. The integrand has poles at $k_{x0}=\frac{\pi}{2}+\frac{\pi\tau}{2}+n\pi+m\pi\tau$, where $n,m \in \mathbb{Z}$. The contributions from the two sides of the contour cancel while the contribution from the top is related to $I_3$ by a quasiperiodic shift of theta functions. }
\end{figure}
\begin{figure}[ht]
\includegraphics[scale=0.6]{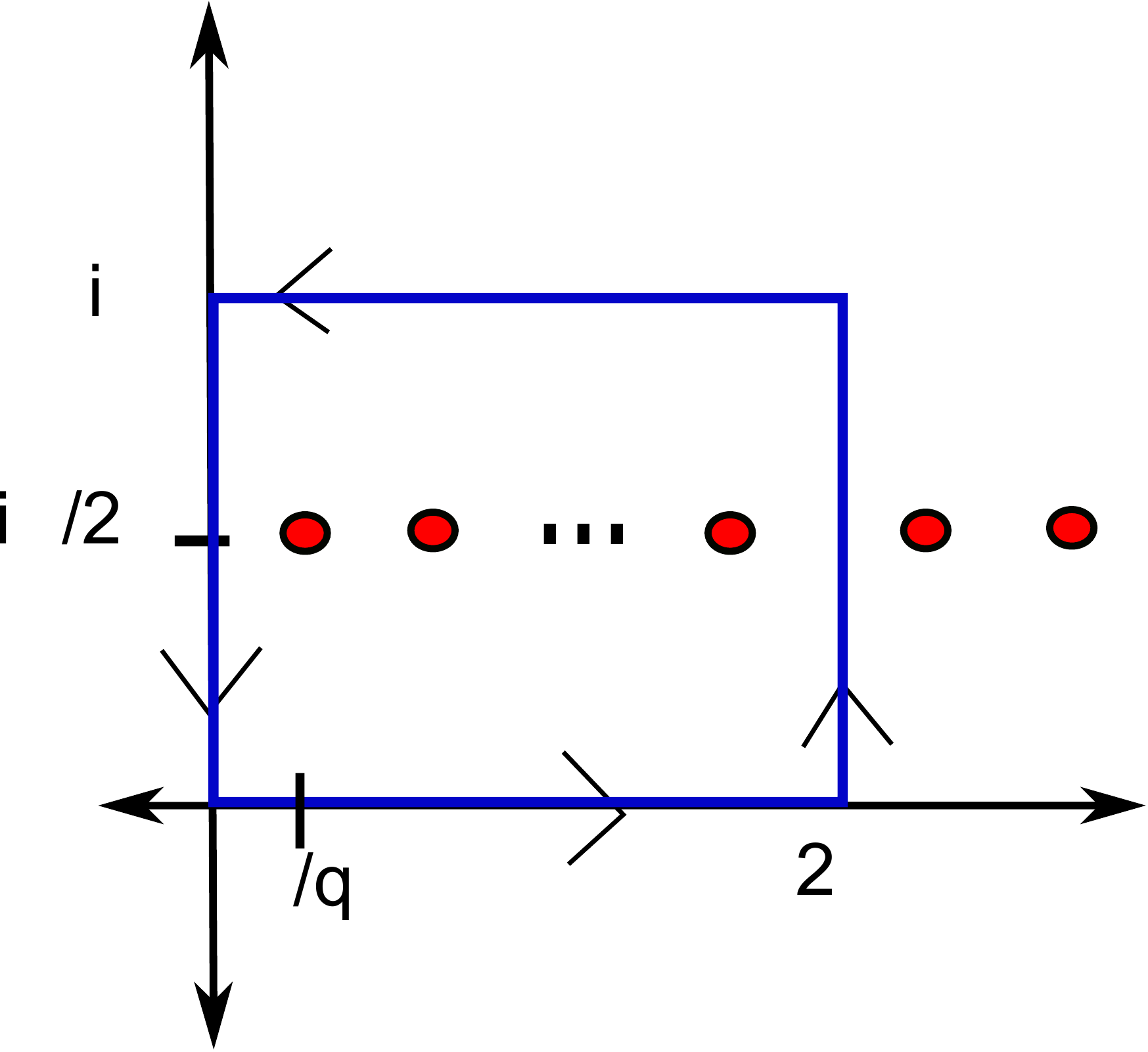}
\caption{\label{fig:I5Contour} (Color online) Contour for the integral $I_5$ in the complex $k_y$ plane. The unit cell shown has $q$ poles in it and the coordinates of the poles are given by $k_{y(m)}=i\frac{\pi}{2}+\frac{\pi}{q}(1+2m)$ where $m=0,1,2....,q-1$. Only the relevant row of poles are shown in the figure.}
\end{figure}

$I_3$ and $I_4$ are evaluated by considering the contour shown in Fig.\ref{fig:I3Contour}. The integrand has poles at $k_0=\frac{\pi}{2}+\frac{\pi\tau}{2}+n\pi+m\pi\tau$, where $n,m \in \mathbb{Z}$. The contour integral shown in the figure is calculated by evaluating a residue at a single pole as 
\begin{equation}
I_C= i \frac{\theta_3\left(\frac{\pi}{2}+i\frac{q\pi}{4}-i \frac{\pi (x-x')}{2}| i \frac{q}{2}\right)}{\theta'_3 \left(\frac{\pi}{2}+i\frac{q\pi}{4}|i \frac{q}{2}\right)}. 
\end{equation} 
Due to the quasiperidicity of the theta functions, the contour integral is related to $I_3$ as
$I_{c}=I_3\left[1-e^{\pi(x-x')}\right]$. Using Jacobi imaginary transform \cite{thetafunctions} $I_3$ is written as
\begin{equation}
I_3=\frac{1}{\sqrt{2q}} \frac{e^{\frac{\pi (x-x')^2}{2q}}}{\sinh [\frac{\pi(x-x')}{2}]} \frac{\theta_1(\frac{\pi (x-x')}{q} |i \frac{2}{q})}{\theta'_1(0|i \frac{q}{2})}.
\end{equation}
By following the same procedure, one can obtain $I_4$ as
\begin{equation}
I_4=\frac{1}{\sqrt{2q}} \frac{e^{\frac{\pi (x-x')^2}{2q}}}{\cosh [\frac{\pi(x-x')}{2}]} \frac{\theta_4(\frac{\pi (x-x')}{q} |i \frac{2}{q})}{\theta'_1(0|i \frac{q}{2})}.
\end{equation}

$I_5$ is evaluated by a similar method, considering the contour in complex $k_y$ plane shown in \ref{fig:I5Contour}. The integrand has poles at the values $k_{y(0)}=-i\frac{\pi}{2}+i n\pi+\frac{\pi}{q}+m\frac{2\pi}{q}$ where $n,m \in \mathbb{Z}$. Once again quasiperiodicity of theta functions relate the contour integral to $I_5$
\begin{equation} I_C=I_5[1-e^{-i\pi (x-x')} e^{-\pi(y-y')}]. \end{equation}

There are $q$ poles inside the integration contour, thus $I_C=2\pi i \sum_{m=0}^{q-1} Res_{(m)}$ where $Res_{(m)}$ is the residue at $m^{th}$ pole. 
Using a series of transformation identities, the contour integral becomes    
\begin{equation}
I_C=-\frac{i}{\sqrt{2q}} \frac{e^{\frac{i\pi (x+x')}{2}} e^{\frac{\pi (x+x')^2}{2q}}e^{i(y-y')(\frac{i\pi}{2}+\frac{\pi}{q})}}{\theta'_1\left(0|\frac{2i}{q}\right)} S,
\end{equation}
where $S$ is a sum over the labels of the poles, $m$ given by
\begin{equation}
S=\sum_{m=0}^{q-1} (-1)^m e^{\frac{2\pi i m(y-y') }{q}} \theta_3\left(\frac{m\pi}{q}+\frac{\pi(1-x-x')}{2q}|\frac{i}{2q}\right).
\end{equation}
This sum is evaluated easily for each term of the series expansion of the theta function, and a resummation yields
\begin{equation}
S=q e^{-\frac{\pi (Y^2}{2q}} e^{-i\frac{\pi Y}{q}} e^{i\frac{\pi Y(x+x')}{q}} \theta_1\left(\frac{i\pi (Y}{2}+\frac{\pi (x+x')}{2}|\frac{i q}{2}\right),
\end{equation}
with $Y=y-y'$.
By using the final expression of the sum $S$ in the contour integral expression in equation(23), one can obtain the contour integral $I_C$, so the integral $I_5$ as
\begin{equation}
I_5=\frac{e^{\frac{\pi (x+x')^2}{2q}}e^{\frac{i\pi (x+x')(y-y')}{q}}}{2\theta'_1(0|\frac{2i}{q})} {\cal M}_{(x-x')}(y-y'),
\end{equation}  
where ${\cal M}_{(x-x')}(y-y')$ is a function of two integers with the definition
\begin{equation}
{\cal M}_a(b)= \left \{ \begin{array} {r@{\quad : \quad}l} \frac{\theta_1\left(\frac{\pi b}{q}|\frac{2 i}{q}\right)}{\sinh\left(\frac{\pi b}{2}\right)} & a \quad \mathrm{even}, \\
\frac{\theta_4\left(\frac{\pi b}{q}|\frac{2 i}{q}\right)}{\cosh\left(\frac{\pi b}{2}\right)} & a \quad \mathrm{odd} . \end{array} \right.
\end{equation}

Finally $I_6$ is related to $I_5$ as,
\begin{equation}
I_6=I_5 (-1)^{(y-y')},
\end{equation} 
which leads to Eq.\ref{AnalyticHopping}.


\end{document}